\documentclass[usenatbib,useAMS]{mnras}

\usepackage{txfonts,amssymb}
\usepackage{amsfonts}
\usepackage{epsfig}

\newlength{\colwidth}
\title [The eccentric planet in CI Tau]
{ The origin of the eccentricity of the hot Jupiter in CI Tau 
}

\author[G. P. Rosotti et al.]
{G. P. Rosotti$^1$\thanks{E-mail:rosotti@ast.cam.ac.uk}, R. A. Booth$^1$, C. J. Clarke$^1$, J. Teyssandier$^2$, S. Facchini$^3$ and A. J. Mustill$^4$\\
$^{1}$ Institute of Astronomy, University of Cambridge, Madingley Road, Cambridge, CB3 0HA, UK\\
$^{2}$ Department of Applied Mathematics and Theoretical Physics, University of Cambridge, Wilberforce Road, Cambridge, CB3 0WA, UK\\
$^{3}$ Max-Planck-Institut f\"{u}r Extraterrestrische Physik, Giessenbachstrasse 1, 85748 Garching, Germany\\
$^{4}$ Lund Observatory, Department of Astronomy and Theoretical Physics, Lund University, Box 43, SE-221 00 Lund, Sweden  } 

\date{Accepted 2016 September 9. Received 2016 September 9; in original form 2016 June 27}

\pagerange{\pageref{firstpage}--\pageref{lastpage}} \pubyear{2016}

\begin {document}
\def\lta{\mathrel{\spose{\lower 3pt\hbox{$\mathchar"218$}}
     \raise 2.0pt\hbox{$\mathchar"13C$}}}
\def\gta{\mathrel{\spose{\lower 3pt\hbox{$\mathchar"218$}}
     \raise 2.0pt\hbox{$\mathchar"13E$}}}

\def\19{GRS~1915+105}
\label{firstpage}
\maketitle

\begin{abstract}
Following the recent discovery of the first radial velocity planet in a
star still possessing a protoplanetary disc (CI Tau), we examine the 
origin of the planet's  eccentricity (e $\sim 0.3$). We show through long timescale
($10^5$ orbits) simulations that the planetary eccentricity can be pumped
by the disc, even when its local surface density is  well below
the threshold previously derived from
short timescale integrations. 
We show that the disc may be able to  excite
the planet's orbital eccentricity in $<$ a Myr for the system parameters
of CI Tau. We also perform two  planet scattering experiments and show
that alternatively the observed planet may plausibly have acquired
its eccentricity through dynamical scattering
of a migrating lower mass planet, which has either been ejected from the system or swallowed by the central star. In the latter case the present location and eccentricity of the observed planet can be recovered if it was previously
stalled within the disc's 
magnetospheric cavity.
\end{abstract}

\begin{keywords}
accretion, accretion discs -- protoplanetary discs -- stars: pre-main-sequence -- planet-disc interactions -- planets and satellites: dynamical evolution and stability\end{keywords}

\section{Introduction}
 
The recent discovery of a radial velocity planet in the young, disc bearing star CI Tau 
\citep{Johns-Krull2016} offers the first opportunity to test theories for the formation and
early evolution of hot Jupiters in discs. To date, planet discoveries
in discs have derived from direct imaging  (e.g \citealt{Chauvin2004,Chauvin2005,Neuhauser2005,Neuhauser2008,Marois2008,Kraus2012,Sallum2015}) 
due to  
difficulties in applying 
transit detection and radial velocity
methods  in young stars. The presence of discs
evidently rules out transit detections (though candidate transit detections
 have  been obtained
in disc-less young stars: \citealt{vanEyken2012,Ciardi2015,David2016}).
Both the transit and radial velocity  techniques are impeded by the extreme
variability of young stars \citep{Xiao2012,Stauffer2014}; in particular it is difficult to disentangle companion 
induced radial velocity variations
from the quasi-periodic signals produced by   
starspots. In the case of  CI Tau, however,
this effect has been  minimised using  K band 
(where starspot activity is reduced); this has  allowed 
the extraction of a radial velocity periodicity   
 ($9$ days) distinct from the photometric 
period  ($7$ days, plausibly ascribed to stellar rotation).

The planet parameters in CI Tau (P = $9$ days, Msini = $8.1$ M$_{\rm Jup} $) 
place it firmly in the "hot Jupiter" category. In contrast to another hot Jupiter recently found around a T Tauri star \citep{Donati2016}, CI Tau also possesses a massive circumstellar disc of $\sim  37$ M$_{\rm Jup}$  
as deduced from previous mm observations using PdBI (\citealt{Guilloteau2011}; see also \citealt{AndrewsWilliams2007}); if the inclination inferred from the outer disc ($45$ to  $54$ degrees; \citealt{Guilloteau2014}) is
also the inclination of the planet then the measured Msini 
corresponds to  a  mass of  $\sim 10$M$_{\rm Jup}$.

 Given the impossibility of forming giant planets {\it in situ} in close proximity to the host star\footnote{See  \citet{ChiangLaughlin2013,HansenMurray2013} for {\it in situ} formation models for planets considerably less massive than that  in CI Tau.},     there is a long-standing debate about the origin of hot Jupiters: whether they arrive in their present locations during the gas rich phase (by disc mediated migration and/or scattering of planetary embryos) or whether instead by dynamical scattering after the disc has dispersed \citep{Lin1996,RasioFord1996}. The recent discovery in CI Tau provides a key demonstration that in at least one object the former is the case. 

The relatively high eccentricity ($e = 0.3 \pm 0.16$)\footnote{See Figure 5 of \citealt{Johns-Krull2016} for a plot of the distribution of the possible values.} is
however somewhat unexpected in a scenario of purely disc mediated
migration, as discs tend to damp planetary eccentricity \citep{Papaloizou2000,TanakaWard2004}\footnote{ Note that {\it stellar tides} raised in the planet are ineffective in
modifying the eccentricity of a planet with these orbital parameters
on a Myr timescale \citep{BarkerOgilvie2009}}. Although the eccentricity of massive  planets can be  
excited by the disc \citep{Papaloizou2001,Dangelo2006,Bitsch2013}, \citet{Dunhill2013} 
have argued that this requires that the disc surface density in the vicinity of the 
planet falls in a restricted range: we will revisit this conclusion through long
timescale FARGO3D integrations of disc planet systems in Section 2. 
Alternatively, such eccentricities
can be driven by interactions involving multiple planets \citep{Marzari2010,Moeckel2012,Lega2013}; in the case of CI Tau, the absence of another period in the radial velocity data implies that any perturbing giant planet is no longer
in the sub- A.U. region and in Section 3 we explore, through two-planet FARGO3D
simulations and through simply parametrised scattering experiments, whether there are 
orbital histories that can generate significant eccentricity in the observed planet
while also  removing the perturber from the inner disc. 
 
  We emphasise that this paper is mainly concerned with 
the excitation of eccentricity
in  the CI Tau radial velocity planet and we do not present
an exhaustive set of scenarios for the system's prior evolutionary history. 
In Section 4 we discuss whether the planet is likely to have
acquired its eccentricity at its current position  and whether
its present  location -  close to but not at  the radius of
corotation between the disc and the star - is significant.

\section{Disc-driven eccentricity growth}
\label{sec:disc-driven}

\citet{Papaloizou2001} first showed that disc driven eccentricity
growth (long established in the case of stellar binaries: e.g. \citealt{Lubow1991a,Lubow1991b}) can be extended to the regime of massive planets, attributing this growth to an instability launched at the 3:1 outer Lindblad
resonance which excites disc eccentricity. More generally, for gap opening planets, \citet{GoldreichTremaine1980} showed that Lindblad resonances lead to growth of eccentricity while corotation resonances lead to its damping (see also \citealt{GoldreichSari2003,Teyssandier2016}). \citet{Dangelo2006} argued for growth of eccentricity for Jupiter mass planets, because of contributions from several Lindblad resonances that lie near the disc
edges (like the 2:4 and 3:5 resonances, with the 1:3 resonance being unimportant here), while corotation resonances are saturated and cannot damp eccentricity (see also \citealt{Duffell2015}).


 The planet in CI Tau is thus in the regime where previous authors have found that
the disc drives eccentricity; this finding extends to the 3D SPH study of 
\citet{Dunhill2013} who also  proposed  a further criterion for
eccentricity driving in terms of a minimum disc surface density  in the vicinity
of the planet, $\Sigma$.  This can be  expressed via  a dimensionless parameter
$q_{\rm{disc}} = \pi \Sigma a ^2/M_p$ where $a$ and $M_p$ are the planet orbital radius
and mass.
\citet{Dunhill2013} proposed that  eccentricity driving
requires $q_{\rm{disc}} > 0.075$; for lower
disc surface densities, the eccentricity rises modestly over a few hundred orbits  
but then declines again.

 In CI Tau, 
 the value of $q_{\rm{disc}}$ can be estimated from 
the observed  accretion rate on to the star ($\dot M = 3 \times 10^{-8} M_\odot$ yr$^{-1}$, \citealt{McClure2013}) and the 
disc temperature at $\sim 0.1$ A.U. derived from  SED modeling ($\sim 1700$ K, \citealt{AndrewsWilliams2007}): in a steady state $\dot M = 3 \pi \nu \Sigma$ where
$\nu$ is the kinematic viscosity. Adopting the  conventional 
$\alpha$ parametrisation for disc viscosity
\citep{ShakuraSunyaev1973} with $\alpha=10^{-3}$ (towards the lower
end of the range of values found in simulations of the
magnetorotational instability in the inner disc 
  \citep[e.g.,][]{Suzuki2010,Flock2013}), we obtain 
an upper limit $q_{\rm{disc}} < 0.014$. \footnote{Note that this value is more than three orders of
magnitude higher than would be obtained by simply extrapolating the disc surface density
profile inferred from submm imaging  on a scale of $\sim 0.5"$, i.e. $\sim 70$ A.U. \citep{AndrewsWilliams2007,Guilloteau2011}}.

 Our upper limit on the value of $q_{\rm{disc}}$ is at face value  below the
threshold required for  eccentricity driving proposed by \citet{Dunhill2013}.
However, they were only able to pursue their computationally expensive
3D SPH simulations over a relatively short time interval (a few hundred orbits).
Given that their simulations show fair agreement with the 2D grid based simulations
of \citet{Papaloizou2001}, it is of interest to use the
FARGO3D code \citep{Benitez2016} in 2D, exploiting graphic processing units (GPU) which accelerate the code significantly, in order to pursue long timescale integration of the disc/planet
system.

\begin{figure}
\includegraphics[width=\columnwidth]{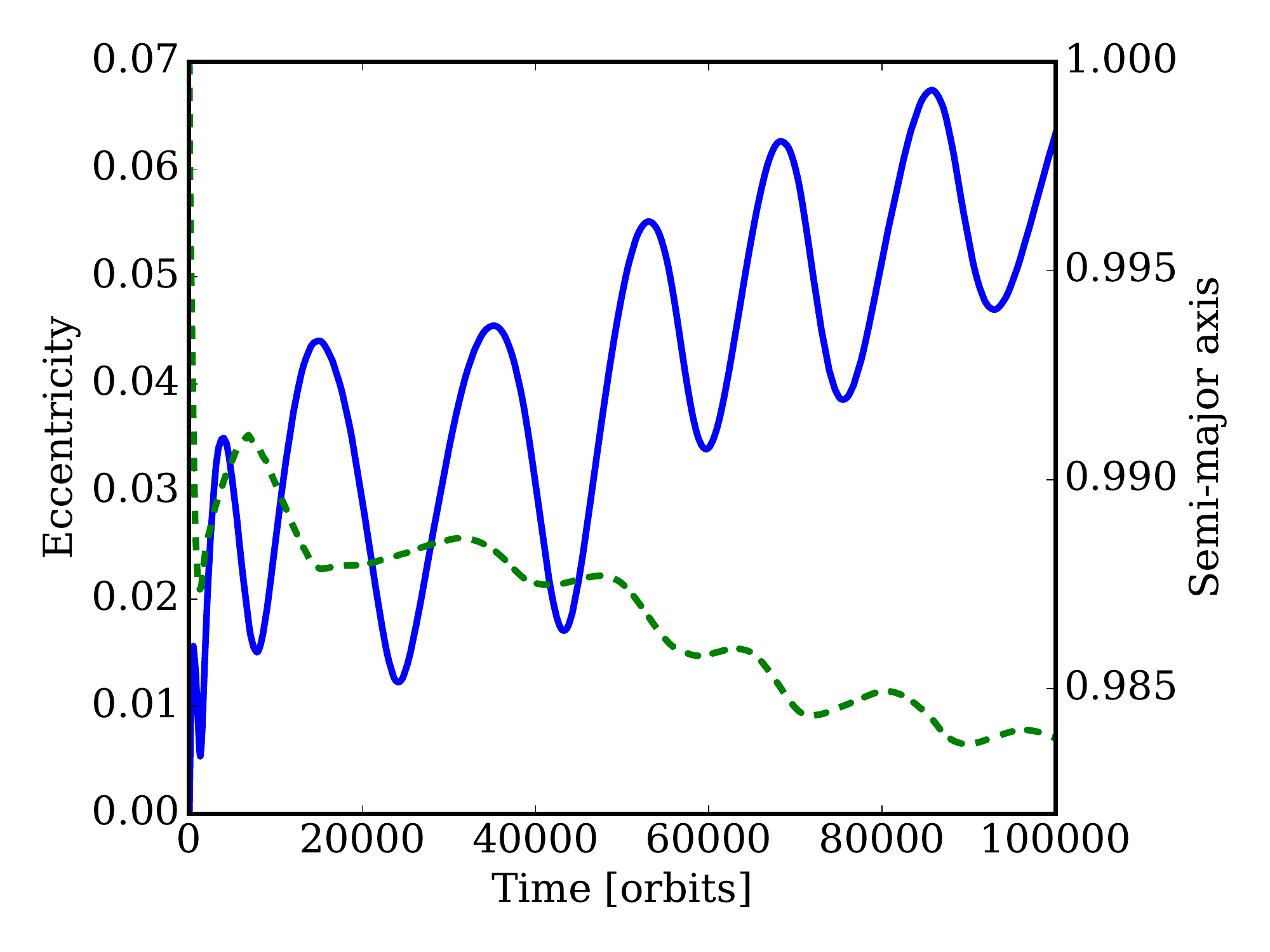}
\caption{Eccentricity (solid blue) and semi-major axis (dashed green) of the planet as a function of time from the 
FARGO3D simulation of a single planet in a disc with $q_\mathrm{disc} = 0.015$. }
\label{fig:eccen_fargo}
\end{figure}

Previous works have either used a ``live'' approach where the planet orbital parameters are  free to evolve, or fixed them and derived their instantaneous rate of change by post processing. Here we choose the former approach, which we motivate later. The simulation is locally isothermal and we assume that the  disc  aspect ratio
varies as  
$0.036 (R/a)^{0.215}$ \citep{AndrewsWilliams2007}. We also adopt  $\alpha =  10^{-3}$ at the location of the planet and a surface density power-law exponent of $0.3$ (as derived from sub-mm imaging), scaling $\alpha$ with radius as $\alpha \propto R^{-0.63}$ in order to 
create a steady state profile 
in the absence of the planet.  
 The numerical grid extends from $0.2$ to $15$  in dimensionless units; the surface density is exponentially tapered, with a tapering radius of 5, to prevent artefacts developing at the outer boundary (since these  were found to affect the results in early tests). The resolution is 430 cells, logarithmically spaced, in radius and 580 in azimuth. The planet has a mass ratio to the central star of 0.013 and is initially fixed on a circular orbit at radius $1$ for the first 50 orbits, over which time its mass is gradually increased up to its final value; thereafter  its orbital parameters evolve freely. The disc surface density normalisation implies $q_\mathrm{disc} = 0.015$, similar to the
upper limit
 derived above.

 We evolve the system over $10^5$ orbits (Figure \ref{fig:eccen_fargo}). 
 On timescales of hundreds of
orbits, the eccentricity evolution is broadly similar to that seen in comparable models in \citet{Dunhill2013}; differences (around a factor three in peak
eccentricity attained) can be readily ascribed to planet mass, disc density profile and viscosity.
In both cases, the eccentricity declines from this first maximum.

 However our long term simulations demonstrate that after $\sim 10^4$  orbits, the eccentricity
begins to grow again and thereafter undergoes oscillatory behaviour superposed on
a slow growth over the duration of the experiment. The oscillatory behaviour
can be understood in terms of secular interaction between the planet and the disc:
the disc develops an eccentric mode which cyclically exchanges eccentricity with the
planetary orbit in a manner reminiscent of  the secular interaction between two eccentric
planets (\citealt{MurrayDermot2000}, Chapter 7). A detailed analysis of this interaction is
postponed to a future paper (Ragusa et al in prep.). Note that this oscillatory interchange
of eccentricity between the planet and the disc can only be captured by a `live'
(freely evolving planet) approach as adopted here. Oscillations appear also in the semi-major axis evolution, which further highlights the need for long-term integrations for studying migration of eccentric planets.

The final eccentricity is already in a regime that overlaps the
broad range of eccentricity values admitted by current orbital solutions of the
planet in CI Tau. More importantly, it is still rising at the end of the simulation;
we estimate that at the current growth rate it will take $5 \times 10^5 - 10^6$ orbits to reach the best fit value of $0.3$, which, given the short orbital timescale at 0.1 A.U., is a small fraction of its current age ($\sim 1$ Myr). The growth rate we measure is slightly lower, but roughly consistent with what was found by \citet{Teyssandier2016} for similar local disc masses in the vicinity of the planet (see their figure 14, although the different setups do not allow for a proper comparison).

 We therefore conclude that our long timescale integrations  provide some
preliminary evidence that the observed eccentricity of the planet in CI Tau may be
the result of pumping by the disc. 

\section{Eccentricity driving by a sibling planet}

 We now consider the alternative scenario in which the eccentricity of the observed
planet is driven by dynamical interaction with another planet. Since
we cannot  explore the parameter space of multiple planet interactions
with
long term hydrodynamical simulations  we adopt
the following approach. We first conducted a single FARGO3D simulation involving
two planets  and disc. We then compared the results with
simple N-body simulations in which the effect of the disc is crudely modelled by 
applying damping of eccentricity and semi-major axis of either or both  planets 
on prescribed timescales, $\tau_e$ and $\tau_a$ (the latter is fixed to $7 \times 10^5 \mathrm{yr}$)\footnote{As discussed in 
Section 2, the disc is expected to drive a slow {\it growth}
of eccentricity for {\it single}  planets with masses above a few $M_{\rm Jup}$. 
However, in resonant {\it multi-planet} systems, eccentricity  is
driven by planet-planet interactions
 and  the disc's role 
is to maintain the two planets close to resonance and to damp the eccentricity
raised by this mutual interaction.}. This comparison allowed us to calibrate the N-body simulations fixing the $\tau_e$/ $\tau_a$ ratio,
which we then used in  N-body calculations with a variety
of planetary configurations. 


\begin{figure}
\includegraphics[width=\columnwidth]{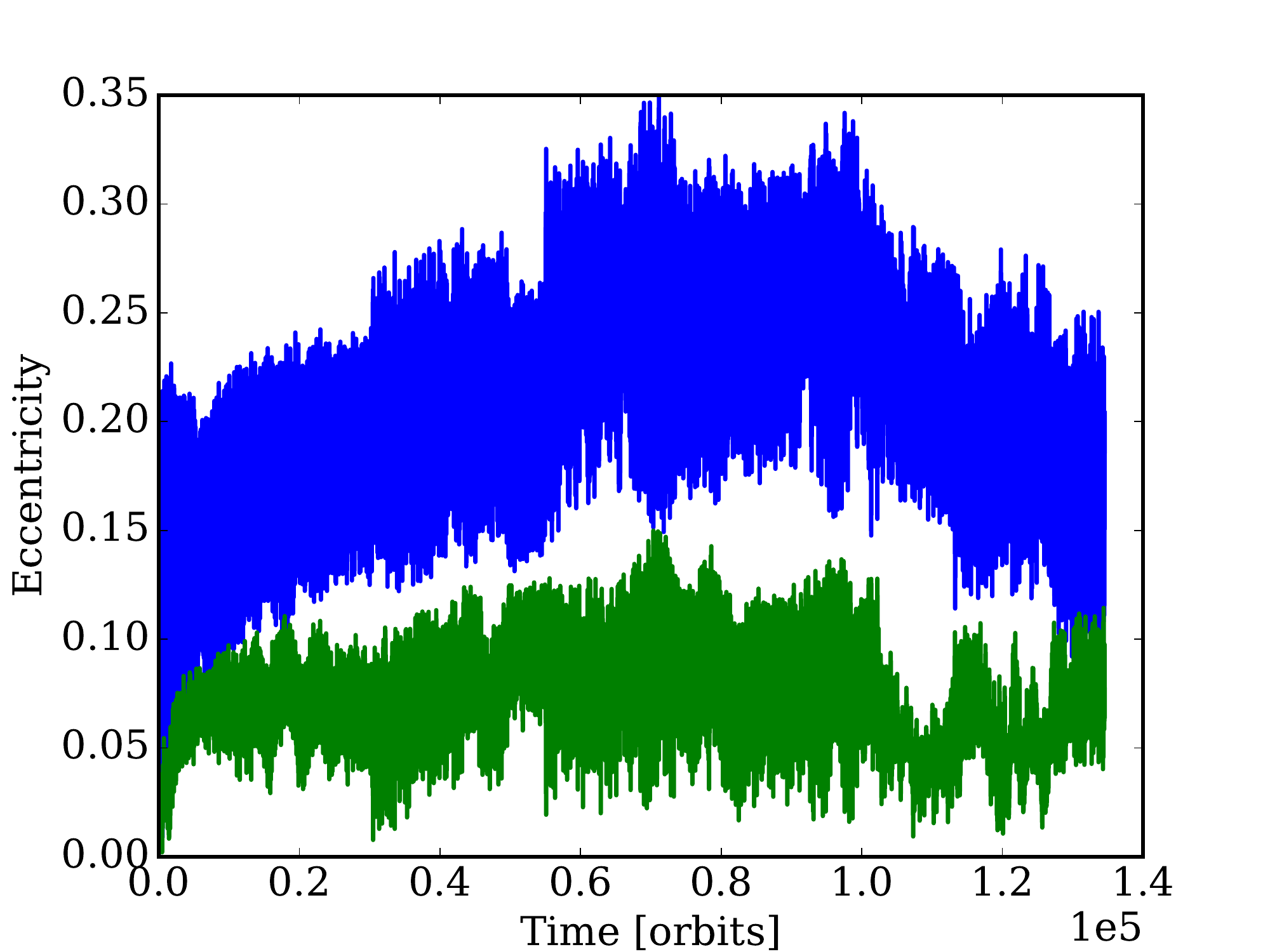}
\caption{Eccentricity evolution of a $10 M_{\rm Jup}$ planet (green) along with an
interior 
$3M_{\rm Jup}$ planet (blue) in a two planet FARGO3D simulation; time is in orbits at $R=1$. See Section 3 for simulation
parameters.}
\label{Fig:Ecc_2Planet}
\end{figure}

 Our hydrodynamical simulations  used  a similar setup to that
described in Section 2,
but  we initially place a $3\,M_{\rm Jup}$ planet at $R=1$ with a $10\,M_{\rm Jup}$ planet
placed outside it in the 2:1 resonance. 
We adopt $q_\mathrm{disc}=0.01$ (normalising to  the mass of the larger planet)
and model the radial  domain up to $R=8$.  We allow the planets to migrate freely under the influence
of the disc and follow their eccentricity evolution for $\sim 10^5$ orbits  
(Fig.~\ref{Fig:Ecc_2Planet}). Within $10^4$ orbits
the planets quickly reach moderate eccentricities, which remain steady (modulo some fluctuations) 
for the full $10^5$ orbits.

\begin{figure*}
\begin{tabular}{cc}
\includegraphics[width=\columnwidth]{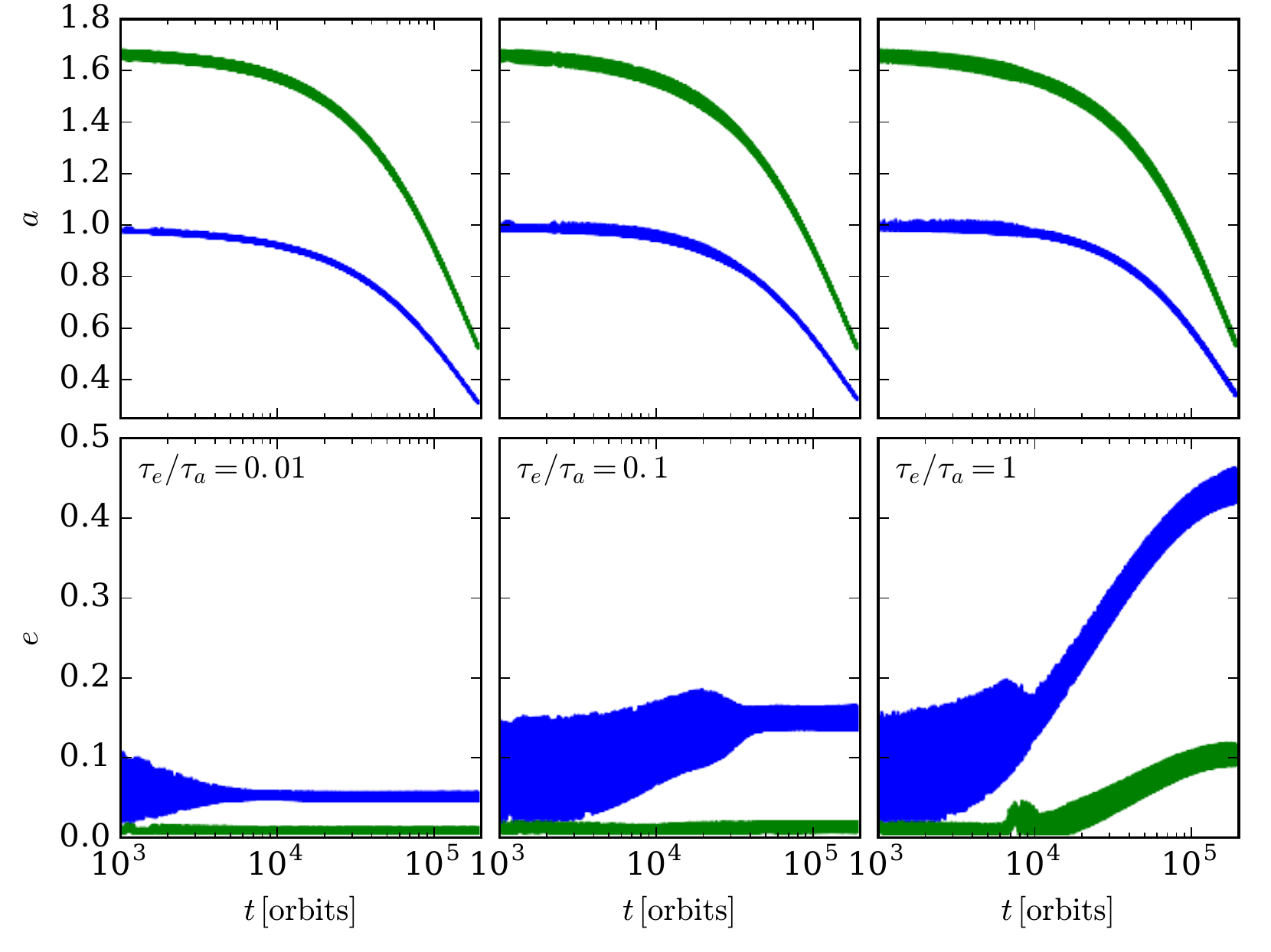} & 
\includegraphics[width=\columnwidth]{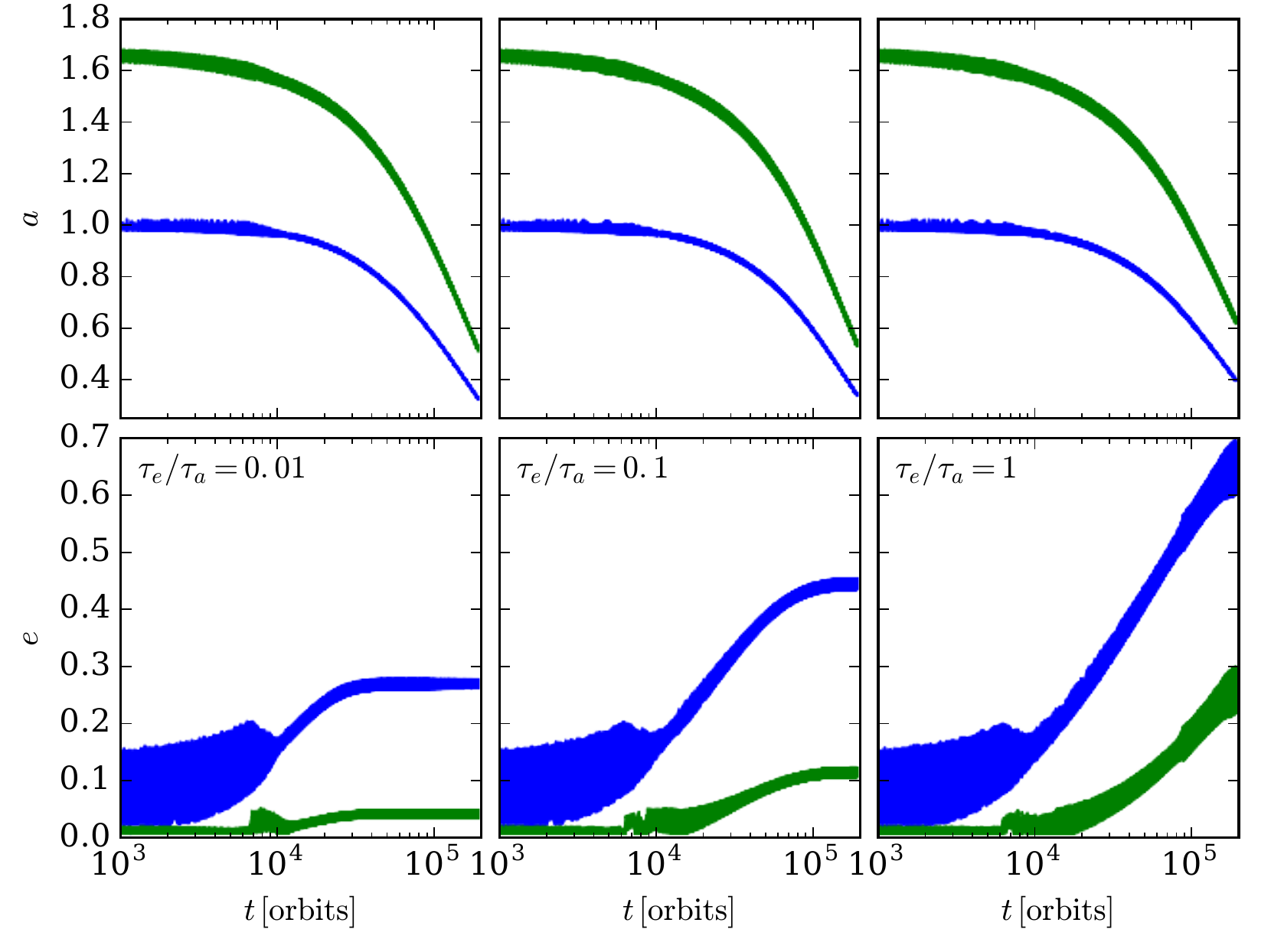} 
\end{tabular}
\caption{Evolution of eccentricity and semi-major axis in two planet  N-body calculations for different ratios of $\tau_e/\tau_a$ (see text). 
A $3\,M_{\rm Jup}$ (blue) and
a $10\,M_{\rm Jup}$ (green) are initially placed at dimensionless radii
$1$ and $1.66$.
Left panel:  eccentricity damping applied to both planets; right panel:
damping only of 
outer planet.
 Time is measured in
orbits at the initial radius of the inner planet.}
\label{Fig:Nbody_evol}
\end{figure*}

\begin{figure}
\includegraphics[width=\columnwidth]{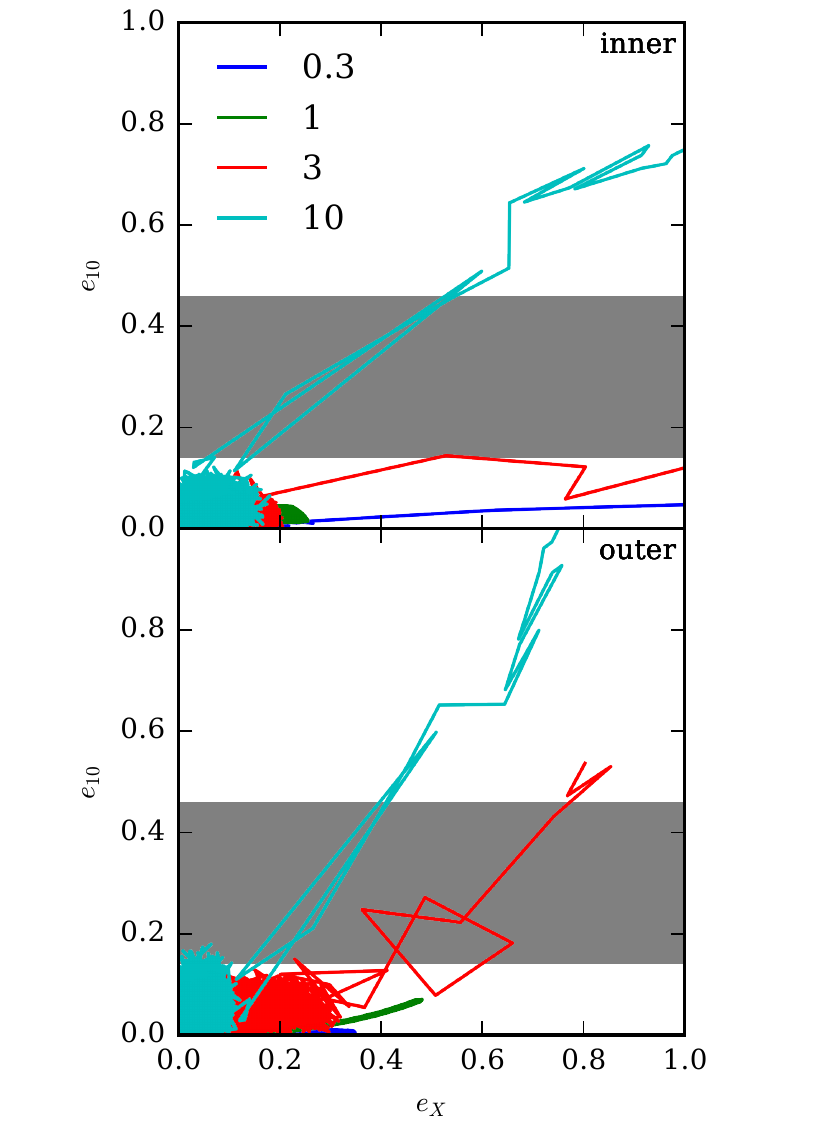}
\caption{Eccentricity of the $10\,M_{\rm Jup}$ planet against the eccentricity of the companion for different companion
masses; in this case $\tau_e/\tau_a=0.1$. The key refers to the mass of the companion in $M_{\rm Jup}$ and inner / outer 
denotes the initial position of the $10\,M_{\rm Jup}$
planet. The lines trace the evolution of the eccentricity from zero, terminating
either at 
 $10^6$ years,  at the ejection of one  of the planets
or at a collision (between the two planets or with 
the star). The grey shaded area shows the $1 \sigma$ interval of the observed eccentricity in CI Tau.}
\label{Fig:e_vs_e}
\end{figure}

The N-body calculations are performed using the code \textsc{mercury} \citep{Chambers1999}, using
the Burlisch-Stoer algorithm. Following \citet{Lee2002} and \citet{Teyssandier2014}, we include additional
damping forces in the N-body model to generate migration and eccentricity 
damping,  
and  vary  the ratio of eccentricity damping to migration time-scales ($\tau_a$ is actually half of the migration time-scale, \citealt{Teyssandier2014})
, $\tau_e / \tau_a$, to match the hydrodynamical  simulation.
The planets (with masses $3\,M_{\rm Jup}$ and  $10\,M_{\rm Jup}$, as in the hydro model) are 
initially given small eccentricities ($\sim 10^{-3}$) and inclinations ($\sim 0.01\deg$).
We consider two scenarios, either  applying damping to both planets or only to the outer planet. 
Neglecting the damping of the inner planet may 
be a good description in the case of
massive
planets which open  a deep gap in the disc and where the inner disc is expected to be depleted.

A comparison of Fig.~\ref{Fig:Ecc_2Planet} with Fig.~\ref{Fig:Nbody_evol} shows that the N-body models can be a fair  match to simulations. When damping on both
planets is included a value of $\tau_e/\tau_a$ in the range $0.1-1$ is needed, while when
the inner planet is undamped we favour values in the range $0.01-0.1$. 
We take  $\tau_e / \tau_a = 0.1$ as a  compromise. We note that our estimate of $\tau_e / \tau_a$
exceeds the typical value of 0.01 estimated from analytical and numerical simulations of single planets with 
$M \lesssim 1\,M_{\rm Jup}$ \citep{Papaloizou2000,Cresswell2007}. This should not be surprising since
we consider planets massive enough to open a deep gap, which means that the co-rotation torques responsible for damping 
the eccentricity are much weaker (which  enables the disc driven eccentricity growth considered in Section 2). The eccentricities attained at the end of the calculations shown in Fig.~\ref{Fig:Nbody_evol} are in broad agreement with the analytical results of \citet{Teyssandier2014}.

We now turn to the question of  whether resonant eccentricity driving can generate  $e \gtrsim 0.2$ as in the observed 
planet in CI Tau while simultaneously removing the sibling planet from a region where it would 
have been detected in the radial velocity data. 
To do this we conducted a suite
of N-body models. We initially place one planet at $0.1\,{\rm au}$ and migrate a second planet into resonance with it (after which they migrate
in resonance together), following the subsequent eccentricity
growth (which can lead to close scattering). In each case we assume that one planet (which may be either the inner or outer planet) has
 mass $10\,M_{\rm Jup}$ 
and vary the mass of the sibling planet.  The only sense in which these calculations are not
scale free is due to the finite size of the star and planets which allows us to distinguish close
scattering events from physical collisions. 

The results of this exercise are shown in Fig.~\ref{Fig:e_vs_e} when only the 
eccentricity of the outer planet is damped. 
The densely filled region of orbital parameters at low
eccentricity corresponds to when the planets are exchanging
eccentricity cyclically but where the eccentricities are sufficiently
low to avoid close scattering events. The
eccentricity of the massive planet never exceeds $0.1-0.2$ during the
former phase; therefore if the eccentricity of the planet in CI Tau
is as high as $\sim 0.3$ then this must have resulted from interactions in the 
latter regime. The scattering regime (typically lasting $\sim 10^3 \mathrm{yr}$) is characterised by large
stochastic excursions in the eccentricity plane  which are terminated
at the point that the planets collide, the lower mass planet is
ejected from the system or the lower mass planet collides with the star
(in the case that the massive planet is the interior one, these outcomes
occur respectively $\sim 5 \%, 55 \%, 40 \%$ of the time in the 100 simulations we have run).
Any of these outcomes would remove the light  planet from the
vicinity of the  heavy planet. We therefore conclude that
in any scenario in which a planet acquired significant eccentricity 
through pumping by a  sibling planet, this sibling should 
{\it not} remain in its vicinity.   
In order to check the sensitivity to our calibration of the eccentricity damping time-scale, we have run models with $\tau_e / \tau_a = 0.01$ or $1$ coming to the same conclusion
--  the eccentricity of the planet in CI Tau cannot have
reached $e > 0.2$ through planet-planet interactions unless a strong
scattering event occurs, and this typically removes the lighter planet from the system.

\section{Discussion \& Conclusion}

 So far we have considered various eccentricity driving mechanisms  
assuming these operate close to the planet's
present position. Given that it is unlikely that the planet formed at
$0.1$ A.U., we need also to consider scenarios for its inward migration from larger
radius.

While there have been conflicting results as to
whether the development of 
eccentricity inhibits disc mediated migration 
\citep{Papaloizou2001,Dangelo2006,Duffell2015,Rice2008}, 
there is some suggestion that eccentricity growth is favoured
at low $H/R$ \citep{Armitage2005}. Since discs are generally flared
($H/R$ increases with $R$), this would favour growth at small radii, possibly 
delaying the excitation of  eccentricity 
until the planet arrives at its present location.
Also, our own simulations, which  
are the only ones to have studied the migration of eccentric
planets  over such long timescales, however find that eccentric
planets {\it can} migrate (see Figure \ref{fig:eccen_fargo})
and in this case the eccentricity may be
excited anywhere between its birthplace and current location. We find
(in simulations for which $e$ is as large as $0.15$) that the planet migrates  
at a rate  that is consistent with the
rate found in simulations involving planets on circular orbits 
(\citet{Duffell2014} and \citet{Duermann2015}). The migration time can be approximated
as $t_{mig} = t_\nu \max(1,\frac{M_p}{\Sigma \pi a^2})$,
where $t_\nu$ is the viscous timescale \footnote{ The formula ignores the recent findings that type II migration is a factor of 2-3 faster than the viscous time-scale as the ones presented here are order of magnitude estimates, but it does catch the correct dependence with the disc mass.}. For  the massive planet considered here, the second
term in brackets is relevant  within $\sim 10$ A.U. (i.e. the planetary inertia is important)
so that
for a steady state disc we have $t_{mig} \sim 10 M_p/\dot M$, independent
of radius and viscosity assumptions. 
For the parameters of CI Tau, this implies $t_{mig} \sim$ a Myr.

 It is tempting to ascribe some significance to the fact that the
radius of the observed planet is
only $\sim 30 \%$ beyond the 
corotation radius between the star and the disc (assuming that
the $7$ day photometric
period of the star measured by  \citealt{Johns-Krull2016} is the star's rotation period). Models
of disc braking of young stars suggest that systems evolve
to a state of disc locking where the disc is truncated slightly
inside the corotation radius. A migrating planet is expected to
stall as it enters the magnetospheric cavity \citep{Romanova2006,Papaloizou2007} so that if
another planet arrives at small radii through disc migration,
a dynamical interaction between the two is assured at this
location. In around $40 \%$ of the cases studied in Section 3,
the remaining $10$ Jupiter mass planet is scattered outward
in the interaction while the lighter sibling planet
ends up being swallowed by the star. This provides a plausible
explanation for why the planet is located in the vicinity of, but
not exactly at, the expected radius of the magnetospheric
cavity. 

Alternatively, if disc mediated migration is effective at moderate eccentricity, then the dynamical
interaction can have occurred at a range of radii. While the subsequent migration
may be accompanied by a damping of the eccentricity to the equilibrium value excited by the disc,
this would take $\sim 10^6\,\mathrm{yr}$ (longer if $\tau_e$ is larger at large $e$,
as found by \citealt{Papaloizou2000}). In this case the proximity of the observed planet to the putative magnetospheric
cavity is coincidental and the planet is likely still migrating.

\section*{Acknowledgements}
We thank J. Papaloizou, L. Prato and E. Ragusa for interesting discussions and the referee for improving the clarity of the paper. This work has been supported by the DISCSIM project, grant agreement 341137 funded by the European Research Council under ERC-2013-ADG, and from STFC through grant ST/L000636/1. This work used the DIRAC Shared Memory Processing and Data Analytic systems, both at the University of Cambridge and operated respectively by the COSMOS Project at the Department of Applied Mathematics and Theoretical Physics and the Cambridge High Performance Computing Service, on behalf of the STFC DiRAC HPC Facility (www.dirac.ac.uk). This equipment was funded by BIS National E-infrastructure capital grants ST/J005673/1 and ST/K001590/1, STFC capital grants ST/H008586/1, ST/H008861/1 and ST/H00887X/1, STFC DiRAC Operations grant ST/K00333X/1, and STFC DiRAC Operations grant ST/K00333X/1. DiRAC is part of the National E-Infrastructure.


\bibliographystyle{mnras}

\begin{thebibliography}{}
\makeatletter
\relax
\def\mn@urlcharsother{\let\do\@makeother \do\$\do\&\do\#\do\^\do\_\do\%\do\~}
\def\mn@doi{\begingroup\mn@urlcharsother \@ifnextchar [ {\mn@doi@}
  {\mn@doi@[]}}
\def\mn@doi@[#1]#2{\def\@tempa{#1}\ifx\@tempa\@empty \href
  {http://dx.doi.org/#2} {doi:#2}\else \href {http://dx.doi.org/#2} {#1}\fi
  \endgroup}
\def\mn@eprint#1#2{\mn@eprint@#1:#2::\@nil}
\def\mn@eprint@arXiv#1{\href {http://arxiv.org/abs/#1} {{\tt arXiv:#1}}}
\def\mn@eprint@dblp#1{\href {http://dblp.uni-trier.de/rec/bibtex/#1.xml}
  {dblp:#1}}
\def\mn@eprint@#1:#2:#3:#4\@nil{\def\@tempa {#1}\def\@tempb {#2}\def\@tempc
  {#3}\ifx \@tempc \@empty \let \@tempc \@tempb \let \@tempb \@tempa \fi \ifx
  \@tempb \@empty \def\@tempb {arXiv}\fi \@ifundefined
  {mn@eprint@\@tempb}{\@tempb:\@tempc}{\expandafter \expandafter \csname
  mn@eprint@\@tempb\endcsname \expandafter{\@tempc}}}

\bibitem[Andrews \& Williams(2007)]{AndrewsWilliams2007} Andrews, S.~M., \& Williams, J.~P.\ 2007, ApJ, 659, 705
\bibitem[Armitage \& Natarajan(2005)]{Armitage2005} Armitage, P.~J., \& Natarajan, P.\ 2005, ApJ, 634, 921 
\bibitem[Barker \& Ogilvie(2009)]{BarkerOgilvie2009} Barker, A.~J., \& Ogilvie, G.~I.\ 2009, \mnras, 395, 2268 
\bibitem[Ben{\'{\i}}tez-Llambay et al.(2016)]{Benitez2016} Ben{\'{\i}}tez-Llambay, P., Ramos, X.~S., Beaug{\'e}, C., \& Masset, F.~S.\ 2016, \apj, 826, 13 
\bibitem[Bitsch et al.(2013)]{Bitsch2013} Bitsch, B., Crida, A., Libert, A.-S., \& Lega, E.\ 2013, A\&A, 555, A124 
\bibitem[\protect\citeauthoryear{{Chambers}}{{Chambers}}{1999}]{Chambers1999} {Chambers} J.~E.,  1999, MNRAS, 304, 793
\bibitem[Chauvin et al.(2004)]{Chauvin2004} Chauvin, G., Lagrange, A.-M., Dumas, C., et al.\ 2004, A\&A, 425, L29 
\bibitem[Chauvin et al.(2005)]{Chauvin2005} Chauvin, G., Lagrange, A.-M., Dumas, C., et al.\ 2005, A\&A, 438, L25 
\bibitem[Chiang \& Laughlin(2013)]{ChiangLaughlin2013} Chiang, E., \& Laughlin, G.\ 2013, MNRAS, 431, 3444 
\bibitem[Ciardi et al.(2015)]{Ciardi2015} Ciardi, D.~R., van Eyken, J.~C., Barnes, J.~W., et al.\ 2015, ApJ, 809, 42 
\bibitem[Cresswell et al.(2007)]{Cresswell2007} Cresswell, P., Dirksen, G., Kley, W., \& Nelson, R.~P.\ 2007, A\&A, 473, 329 
\bibitem[D'Angelo et al.(2006)]{Dangelo2006} D'Angelo, G., Lubow, S.~H., \& Bate, M.~R.\ 2006, ApJ, 652, 1698
\bibitem[David et al.(2016)]{David2016} David, T., Hillenbrand, L., Petigura, E., et al.\ 2016, \nat, 534, 658 
\bibitem[Donati et al.(2016)]{Donati2016} Donati, J., Moutou, C, Malo, L, et al.\ 2016, arXiv:1606.06236
\bibitem[D{\"u}rmann \& Kley(2015)]{Duermann2015} D{\"u}rmann, C., \& Kley, W.\ 2015, A\&A, 574, A52
\bibitem[Duffell et al.(2014)]{Duffell2014} Duffell, P.~C., Haiman, Z., MacFadyen, A.~I., D'Orazio, D.~J., \& Farris, B.~D.\ 2014, ApJ, 792, L10 
\bibitem[Duffell \& Chiang(2015)]{Duffell2015} Duffell, P.~C., \& Chiang, E.\ 2015, ApJ, 812, 94 
\bibitem[Dunhill \& Alexander(2013)]{Dunhill2013} Dunhill, A.~C., \& Alexander, R.~D.\ 2013, MNRAS, 435, 2328
\bibitem[Flock et al.(2013)]{Flock2013} Flock, M., Fromang, S., Gonz{\'a}lez, M., \& Commer{\c c}on, B.\ 2013, A\&A, 560, A43 
\bibitem[Kraus \& Ireland(2012)]{Kraus2012} Kraus, A.~L., \& Ireland, M.~J.\ 2012, ApJ, 745, 5 
\bibitem[Goldreich \& Tremaine(1980)]{GoldreichTremaine1980} Goldreich, P., \& Tremaine, S.\ 1980, \apj, 241, 425
\bibitem[Goldreich \& Sari(2003)]{GoldreichSari2003} Goldreich, P., \& Sari, R.\ 2003, \apj, 585, 1024  
\bibitem[Guilloteau et al.(2011)]{Guilloteau2011} Guilloteau, S., Dutrey, A., Pi{\'e}tu, V., \& Boehler, Y.\ 2011, A\&A, 529, A105
\bibitem[Guilloteau et al.(2014)]{Guilloteau2014} Guilloteau, S., Simon, M., Pi{\'e}tu, V., et al.\ 2014, A\&A, 567, A117 
\bibitem[Johns-Krull et al.(2016)]{Johns-Krull2016} Johns-Krull, C.~M., McLane, J.~N., Prato, L., et al.\ 2016, arXiv:1605.07917 
\bibitem[Hansen \& Murray(2013)]{HansenMurray2013} Hansen, B.~M.~S., \& Murray, N.\ 2013, ApJ, 775, 53 
\bibitem[Lee \& Peale(2002)]{Lee2002} Lee, M.~H., \& Peale, S.~J.\ 2002, ApJ, 567, 596 
\bibitem[Lega et al.(2013)]{Lega2013} Lega, E., Morbidelli, A., \& Nesvorn{\'y}, D.\ 2013, MNRAS, 431, 3494 
\bibitem[Lin et al.(1996)]{Lin1996} Lin, D.~N.~C., Bodenheimer, P., \& Richardson, D.~C.\ 1996, Nature, 380, 606 
\bibitem[Lubow(1991a)]{Lubow1991a} Lubow, S.~H.\ 1991, ApJ, 381, 259 
\bibitem[Lubow(1991b)]{Lubow1991b} Lubow, S.~H.\ 1991, ApJ, 381, 268
\bibitem[Kley \& Dirksen(2006)]{Kley2006} Kley, W., \& Dirksen, G.\ 2006, A\&A, 447, 369 
\bibitem[Marois et al.(2008)]{Marois2008} Marois, C., Macintosh, B., Barman, T., et al.\ 2008, Science, 322, 1348 
\bibitem[Marzari et al.(2010)]{Marzari2010} Marzari, F., Baruteau, C., \& Scholl, H.\ 2010, A\&A, 514, L4 
\bibitem[McClure et al.(2013)]{McClure2013} McClure, M.~K., D'Alessio, P., Calvet, N., et al.\ 2013, ApJ, 775, 114 
\bibitem[Moeckel \& Armitage(2012)]{Moeckel2012} Moeckel, N., \& Armitage, P.~J.\ 2012, MNRAS, 419, 366 
\bibitem[Murray \& Dermott(2000)]{MurrayDermot2000} Murray, C.~D., \& Dermott, S.~F.\ 2000, Solar System Dynamics, Cambridge University Press, Cambridge, UK
\bibitem[Neuh{\"a}user et al.(2005)]{Neuhauser2005} Neuh{\"a}user, R., Guenther, E.~W., Wuchterl, G., et al.\ 2005, A\&A, 435, L13 
\bibitem[Neuh{\"a}user et al.(2008)]{Neuhauser2008} Neuh{\"a}user, R., Mugrauer, M., Seifahrt, A., Schmidt, T.~O.~B., \& Vogt, N.\ 2008, A\&A, 484, 281
\bibitem[\protect\citeauthoryear{{Papaloizou} \& {Larwood}}{{Papaloizou} \& {Larwood}}{2000}]{Papaloizou2000} {Papaloizou} J.~C.~B.,  {Larwood} J.~D.,  2000, MNRAS, 315, 823
\bibitem[Papaloizou et al.(2001)]{Papaloizou2001} Papaloizou, J.~C.~B., Nelson, R.~P., \& Masset, F.\ 2001, A\&A, 366, 263 
\bibitem[Papaloizou(2007)]{Papaloizou2007} Papaloizou, J.~C.~B.\ 2007, A\&A, 463, 775 
\bibitem[Rasio \& Ford(1996)]{RasioFord1996} Rasio, F.~A., \& Ford, E.~B.\ 1996, Science, 274, 954 
\bibitem[Rice et al.(2008)]{Rice2008} Rice, W.~K.~M., Armitage, P.~J., \& Hogg, D.~F.\ 2008, MNRAS, 384, 1242 
\bibitem[Romanova \& Lovelace(2006)]{Romanova2006} Romanova, M.~M., \& Lovelace, R.~V.~E.\ 2006, ApJ, 645, L73 
\bibitem[Sallum et al.(2015)]{Sallum2015} Sallum, S., Follette, K.~B., Eisner, J.~A., et al.\ 2015, Nature, 527, 342 
\bibitem[Shakura \& Sunyaev(1973)]{ShakuraSunyaev1973} Shakura, N.~I., \& Sunyaev, R.~A.\ 1973, A\&A, 24, 337
\bibitem[Suzuki et al.(2010)]{Suzuki2010} Suzuki, T.~K., Muto, T., \& Inutsuka, S.-i.\ 2010, ApJ, 718, 1289 
\bibitem[Stauffer et al.(2014)]{Stauffer2014} Stauffer, J., Cody, A.~M., Baglin, A., et al.\ 2014, AJ, 147, 83 
\bibitem[Tanaka \& Ward(2004)]{TanakaWard2004} Tanaka, H., \& Ward, W.~R.\ 2004, ApJ, 602, 388
\bibitem[Teyssandier \& Ogilvie(2016)]{Teyssandier2016} Teyssandier, J., \& Ogilvie, G.~I.\ 2016, MNRAS, 458, 3221 
\bibitem[Teyssandier \& Terquem(2014)]{Teyssandier2014} Teyssandier, J., \& Terquem, C.\ 2014, MNRAS, 443, 568 
\bibitem[van Eyken et al.(2012)]{vanEyken2012} van Eyken, J.~C., Ciardi, D.~R., von Braun, K., et al.\ 2012, ApJ, 755, 42 
\bibitem[Xiao et al.(2012)]{Xiao2012} Xiao, H.~Y., Covey, K.~R., Rebull, L., et al.\ 2012, ApJS, 202, 7 

\makeatother
\label{lastpage}

\end{thebibliography}

{}
\end{document}